\documentclass[journal,10pt]{IEEEtran}
\normalsize
\usepackage{cite}
\usepackage{amsmath,amssymb,amsfonts}
\usepackage{algorithmic}
\usepackage{graphicx,color}
\usepackage{textcomp}

\usepackage[dvipsnames]{xcolor}
\usepackage{dsfont}
\usepackage{diagbox}
\usepackage{gensymb}
\usepackage{lineno}
\usepackage{amsthm}
\usepackage[normalem]{ulem}
\usepackage{balance}
\usepackage{epsfig}
\usepackage{bm}
\usepackage{enumerate}
\usepackage{latexsym}
\usepackage{graphics}
\usepackage{subfigure}
\usepackage{cite}
\usepackage{mathrsfs}
\usepackage{stfloats}
\usepackage{extarrows}
\usepackage{upgreek}
\usepackage{lineno}
\usepackage{caption}
\usepackage{multirow}
\usepackage[linesnumbered,ruled,vlined]{algorithm2e}

\theoremstyle{definition}

\newtheorem{lemma}{Lemma}

\newtheorem{remark}{Remark}

\newcommand{\vect}[1]{\mathbf{#1}}

\def\j{\mathrm{j}}
\def\tr{\mathrm{tr}}

\definecolor{c1}{HTML}{bc6657}
\definecolor{c2}{HTML}{a76a36}

\begin{document}
\title{Beamfocusing and Power Allocation for AN-Based PLS in Multiuser XL-MIMO with Multiple Eavesdroppers}
\author{
Xiangjun Ma, Ali Arshad Nasir, and Daniel Benevides da Costa
\thanks{
%
Xiangjun Ma, Ali Arshad Nasir, and Daniel Benevides da Costa are with the Interdisciplinary Research Center for Communication Systems and Sensing, Department of Electrical Engineering, King Fahd University of Petroleum $\&$ Minerals, Dhahran 31261, Saudi Arabia (e-mails: ezio1390265506@hotmail.com; anasir@kfupm.edu.sa; danielbcosta@ieee.org).
}\vspace{-0.1cm}
}
\maketitle

\begin{abstract}
This paper investigates the downlink (DL) physical layer security (PLS) in a near-field (NF) extra-large multiple-input multiple-output MIMO (XL-MIMO) system. To enhance the secrecy rate (SR), null-space artificial noise (AN) is transmitted alongside the confidential message, ensuring orthogonality with legitimate user equipment (LUE) channels.
The objective is to maximize the minimum SR by optimizing the NF beamfocusing matrix and power allocation between the signal and AN, considering various channel state information (CSI) conditions and transmit power constraints.
The proposed approach uses successive convex approximation (SCA) for beamfocusing optimization and golden section search (GSS) for power allocation.
The following open questions are addressed: (i) Can AN transmission further enhance SR for multiple LUEs in the presence of multiple eavesdropping user equipment (EUEs)? (ii) Can null-space AN transmission achieve attractive SR performance even without CSI availability for EUEs? Both questions are affirmatively answered and explored in detail, with an algorithm presented for joint beamfocusing design and AN-aided power allocation. The proposed method outperforms state-of-the-art approaches that either omit AN transmission or rely on maximal-ratio transmission (MRT) for beamfocusing.
\end{abstract}

\maketitle

\section{Introduction} \label{Section:Introduction}

As a promising technique in the upcoming sixth-generation (6G) networks, extremely large-scale multiple-input multiple-output (XL-MIMO) systems are equipped with a large number of
antennas, resulting in an oversized antenna aperture.
This significantly improves spectral efficiency (SE) compared to conventional MIMO systems \cite{9903389}.
On the other hand, the electromagnetic (EM) radiation from the transmitter to the receiver can be divided into far-field (FF) and near-field (NF) regions, differentiated by the Rayleigh distance.
Since the Rayleigh distance is positively correlated with the antenna aperture, NF communications become dominant in XL-MIMO systems. This necessitates the use of spherical-wave model to accurately characterize the channel.

Due to the broadcast nature of the wireless propagation channel, transmitted signals from base station (BS) to legitimate UEs (LUEs) are vulnerable to wiretapping by malicious eavesdropping UEs (EUEs).
For FF communication, the beamforming design based on the planar-wave model often leads to significant leakage of confidential messages, as EUEs are typically located in the same direction as LUEs \cite{10436390}.
In contrast, the spherical-wave model accounts for both the angle and distance information of the receiver, allowing precoding to focus on a specific region, known as beamfocusing.
This approach enables positive secrecy rate (SR) even when EUEs are positioned in the same direction.

Several groundbreaking studies have recently emerged on physical layer security (PLS) for NF communications.
For the single-UE scenario, the authors in \cite{9860861} investigated secrecy performance of XL-MIMO, where simulation results demonstrated that the spherical-wave model can achieve higher SR.
An NF secure transmission framework using hybrid beamforming was proposed in \cite{10436390}, where a two-stage algorithm was developed to maximize SR.
The authors in \cite{10504668} investigated the SR maximization problem in NF wideband communications, where a novel analog beamfocusing scheme was introduced to address the interplay between near-field propagation and wideband beamsplit.
For the multi-UE scenario, \cite{9967979} investigated the sum SR in XL-MIMO and proposed a novel leakage subspace precoding strategy, demonstrating superior performance compared to conventional methods.
In \cite{10540207}, a hybrid beamfocusing scheme was introduced to maximize the minimum SR in NF communications, ensuring a more uniform SR distribution among all LUEs.

Since the passive EUE is undetectable, its channel state information CSI is hard to estimate. A practical approach to enhance secrecy is injecting artificial noise (AN).
In \cite{9133130}, the PLS of a reconfigurable intelligent surface (RIS)-aided MIMO system was investigated, where the beamforming matrix and AN covariance matrix were jointly optimized to maximize the sum-rate while limiting confidential message leakage to EUEs.
The work in \cite{10552246} examined the secrecy performance of a RIS-aided cell-free network, formulating a max-min security energy efficiency optimization problem to jointly design active beamforming and AN matrices at BSs and passive beamforming matrices at RISs.
However, both studies \cite{9133130} and \cite{10552246} focused on conventional FF communication. A novel AN-aided beamfocusing scheme was developed in \cite{Zhifeng2025Low} for a NF communication system.
However, the study did not address the challenging max-min SR optimization in the presence of multiple LUE-EUE case.
Additionally, the adopted maximal-ratio transmission (MRT) beamfocusing usually does not yield optimal SR and the assumption of perfect CSI of EUEs is not practical as well.

Building on the preceding discussion, this paper investigates secure downlink (DL) transmission in an NF XL-MIMO system, where both LUEs and EUEs are aligned in the same direction relative to the BS. To enhance the SR, null-space AN is transmitted alongside the signal, ensuring orthogonality to LUE channels.
The objective is to maximize the minimum SR by optimizing the NF beamfocusing matrix and the power allocation between the signal and AN, considering various CSI conditions and subject to the BS's power budget.
The proposed approach leverages successive convex approximation (SCA) for beamfocusing optimization and golden section search (GSS) for power allocation.
We have demonstrated the superiority of the proposed algorithm over existing state-of-the-art approaches that either omit AN transmission or rely on MRT for beamfocusing. Our simulation results show that the proposed null-space AN transmission, under no CSI availability of EUEs, can achieve a performance close to the upper benchmark, which assumes CSI availability of EUEs.


\emph{Notations}:
The superscripts $^\mathrm{T}$ and $^\mathrm{H}$ denote the and Hermitian transpose, respectively. $\|\bullet\|$ and $\|\bullet\|_\text{F}$ denote the 2-norm and Frobenius norm, respectively.
$\vect{I}_N$ is the $N\times{N}$ identity matrix.
The notation $[\vect{a}]_n$ refers to the $n$-th element of vector $\vect{a}$.
The indicator function $\mathds{1}_{\mathcal{P}}(x)$ is defined as
$\mathds{1}_{\mathcal{P}}(x)=1$ if $x\in\mathcal{P}$ and $\mathds{1}_{\mathcal{P}}(x)=0$ if $x\notin\mathcal{P}$.
Finally, $[x]^+=\max\{0,x\}$.

\section{System Model} \label{Section:system_model}

\begin{figure}
\begin{center}
\includegraphics[width=0.9\columnwidth]{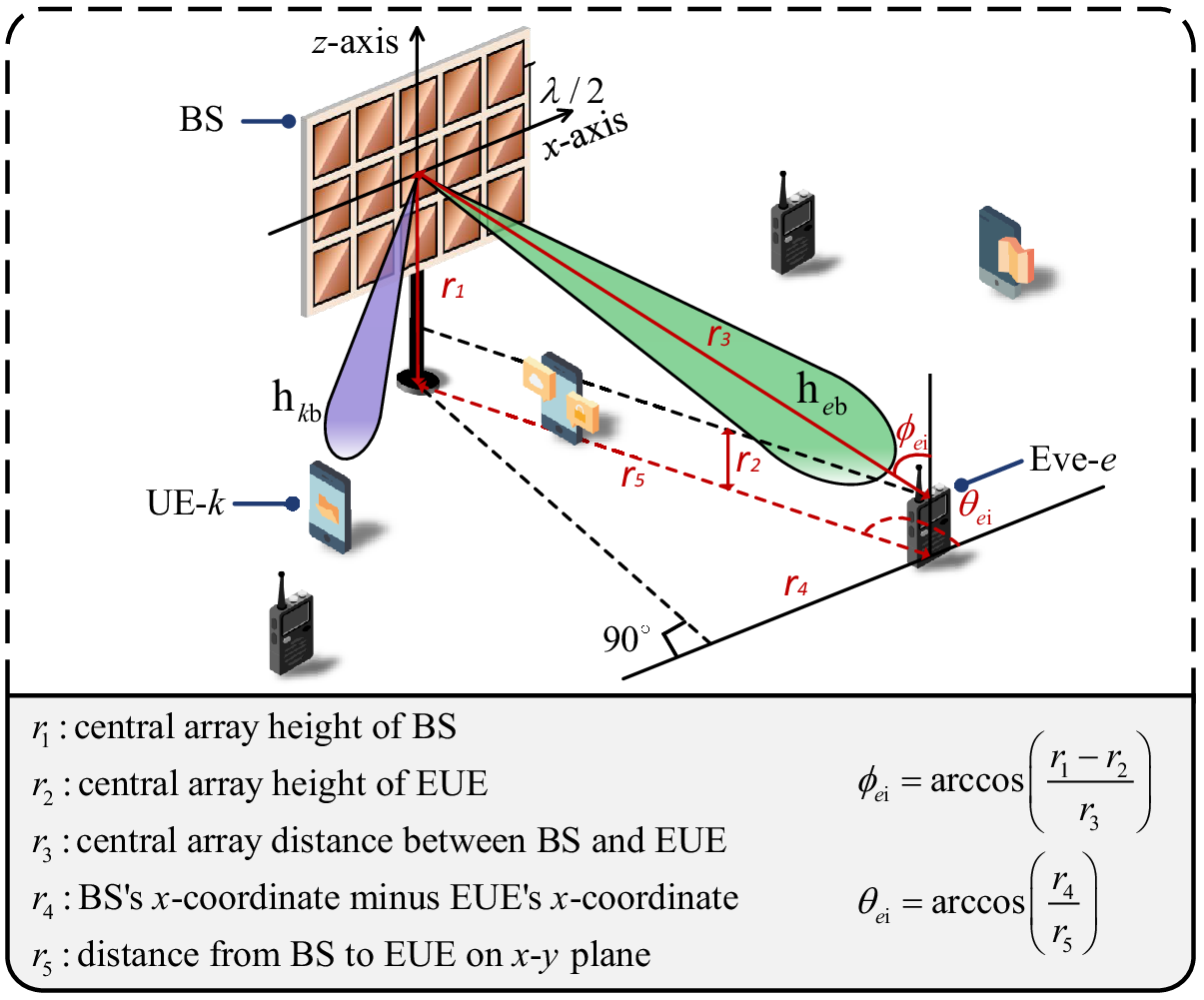} 
\captionsetup{font={normal}}
\caption{DL secrecy transmission of XL-MIMO.}\label{Figure:system_model}
\end{center} \vskip-3mm
\end{figure}

We consider a DL secrecy transmission scenario in an XL-MIMO system, as illustrated in Fig. \ref{Figure:system_model}. In this setup, a BS equipped with $N_\mathrm{b}$ antennas communicates with $K$ single-antenna LUEs, while $E$ single-antenna EUEs attempt to intercept the confidential messages intended for all $K$ LUEs.
We define the LUE and EUE sets as $\mathcal{K}=\{1,\cdots,K\}$ and $\mathcal{E}=\{1,\cdots,E\}$, respectively,
and further define $\mathcal{U}=\mathcal{K}\cup\mathcal{E}$.
The total number of BS antennas is given by $N_\mathrm{b}=N_\mathrm{b}^xN_\mathrm{b}^z$, where
$N_\mathrm{b}^{x}$ and $N_\mathrm{b}^{z}$ represent the number of elements in the uniform planar array (UPA) along the $x$-axis and $z$-axis, respectively.
Moreover, we assume that the number of UPA elements in each direction is odd, i.e.,
$N_{\mathrm{b}}^{\nu}=2\bar{N}_{\mathrm{b}}^{\nu}+1$, where
$\bar{N}_{\mathrm{b}}^{\nu}$ is a non-negative integer and $\nu\in\{x,z\}$.

\subsection{Channel Model}

The inter-antenna distance is set to $\lambda/2$, where $\lambda$ is the wavelength.
The channel from the BS to UE-$u$ ($u\in\mathcal{U}$) is denoted as $\vect{h}_{u\mathrm{b}}^\mathrm{H}\in\mathbb{C}^{1\times{N}_\mathrm{b}}$, and can be modeled as:
\begin{equation}
\vect{h}_{u\mathrm{b}}^\mathrm{H}=\beta_{u\mathrm{b}}^{\frac{1}{2}}e^{-\j\frac{2\pi}{\lambda}d_{u\mathrm{b}}}
[\vect{a}_\mathrm{b}(\theta_{u\mathrm{b}},\phi_{u\mathrm{b}},d_{u\mathrm{b}})]^\mathrm{T}
+\sum_{l=1}^{L}(\vect{h}_{u\mathrm{b}}^{(l)})^\mathrm{H}.
\end{equation}
where $L$ is the number of non-line-of-sight (NLoS) paths. Additionally,
$\beta_{u\mathrm{b}}$, $d_{u\mathrm{b}}$, $\theta_{u\mathrm{b}}$, and $\phi_{u\mathrm{b}}$ represent the LSFC, distance, azimuth angle, and elevation angle of the line-of-sight (LoS) path from the BS to UE-$u$, respectively.
Furthermore, $\vect{a}_\mathrm{b}(\theta_{u\mathrm{b}},\phi_{u\mathrm{b}},d_{u\mathrm{b}})\in\mathbb{C}^{N_\mathrm{b}}$ denotes the transmitting array response of the BS,
and $(\vect{h}_{u\mathrm{b}}^{(l)})^\mathrm{H}$ represents the $l$-th NLoS path, as discussed in \cite{10220205}.
Define
\begin{align}
\notag&{f}^x(\theta,\phi,d,\bar{n})=
\exp\bigg(-\j\frac{2\pi}{\lambda}\bigg(\bar{n}\frac{\lambda}{2}
\cos(\theta)\sin(\phi)
\\&+\frac{\bar{n}^2}{2d}\frac{\lambda^2}{4}(1-\cos^2(\theta)\sin^2(\phi))\bigg)\bigg),
\\&{f}^z(\phi,d,\bar{n})=\exp\bigg(-\j\frac{2\pi}{\lambda}\bigg(\bar{n}\frac{\lambda}{2}\cos(\phi)
+\frac{\bar{n}^2}{2d}\frac{\lambda^2}{4}\sin^2(\phi)\bigg)\bigg).
\end{align}
The array responses can be expressed as
\begin{subequations}
\begin{align}
&{\vect{a}}_\mathrm{b}(\theta_{u\mathrm{b}},\phi_{u\mathrm{b}},d_{u\mathrm{b}})
={\vect{a}}_\mathrm{b}^x(\theta_{u\mathrm{b}},\phi_{u\mathrm{b}},d_{u\mathrm{b}})
\otimes{\vect{a}}_\mathrm{b}^z(\phi_{u\mathrm{b}},d_{u\mathrm{b}}),\label{a_b_negative}
\\&[{\vect{a}}_\mathrm{b}^x(\theta_{u\mathrm{b}},\phi_{u\mathrm{b}},d_{u\mathrm{b}})]_{{n}_\mathrm{b}^x}=
{f}^x(\theta_{u\mathrm{b}},\phi_{u\mathrm{b}},d_{u\mathrm{b}},{\bar{n}}_\mathrm{b}^x),
\\&[{\vect{a}}_\mathrm{b}^z(\phi_{u\mathrm{b}},d_{u\mathrm{b}})]_{{n}_\mathrm{b}^z}=
{f}^z(\phi_{u\mathrm{b}},d_{u\mathrm{b}},{\bar{n}}_\mathrm{b}^z),
\end{align}
\end{subequations}
where ${n}_{\mathrm{b}}^{\nu}={\bar{n}}_{\mathrm{b}}^{\nu}+\bar{N}_{\mathrm{b}}^{\nu}+1$, with $\bar{n}_{\mathrm{b}}^{\nu}\in\{-\bar{N}_{\mathrm{b}}^{\nu},\cdots,\bar{N}_{\mathrm{b}}^{\nu}\}$.

Let $\vect{x}\in\mathbb{C}^{{N}_\mathrm{b}}$ be the precoded transmitting signal of the BS. The received signal at UE-$u$ is given by \cite{7328729}
\begin{subequations}
\begin{align}
&y_u=\vect{h}_{u\mathrm{b}}^\mathrm{H}\vect{x}+\vect{n}_u,
\\&\vect{x}=\sqrt{\epsilon_\mathrm{s}}\vect{W}\vect{s}+\sqrt{\epsilon_\mathrm{a}}\vect{V}\vect{z}_\mathrm{a}.
\end{align}
\end{subequations}
Here, $\vect{s}\sim\mathcal{N}_\mathbb{C}(\vect{0}_K,\vect{I}_K)$ and $\vect{z}_\mathrm{a}\sim\mathcal{N}_\mathbb{C}(\vect{0}_{N_\mathrm{b}},\vect{I}_{N_\mathrm{b}})$ are the data and AN vectors, respectively.
The normalized data and AN precoding matrices are $\vect{W}=[\vect{w}_1,\cdots,\vect{w}_K]\in\mathbb{C}^{{N}_\mathrm{i}\times{K}}$ and $\vect{V}\in\mathbb{C}^{{N}_\mathrm{b}\times{N}_\mathrm{b}}$, with $\tr(\vect{W}^\mathrm{H}\vect{W})=K$ and $\tr(\vect{V}^\mathrm{H}\vect{V})=N_\mathrm{b}$.
Additionally,
$\epsilon_\mathrm{s}={\epsilon{P}_\mathrm{b}}/{K}$ and $\epsilon_\mathrm{a}={(1-\epsilon){P}_\mathrm{b}}/{N_\mathrm{b}}$, where ${P}_\mathrm{b}$ is the transmit power of BS, and $\epsilon\in[0,1]$ is the power allocation factor.
Finally, $\vect{n}_u\sim\mathcal{N}_\mathbb{C}(\vect{0}_{N_u},\sigma^2\vect{I}_{N_u})$ represents the additive white Gaussian noise (AWGN) at UE-$u$.
Let $\vect{H}_{\mathrm{b}}^\mathrm{H}=[\vect{h}_{1\mathrm{b}},\cdots,\vect{h}_{K\mathrm{b}}]^\mathrm{H}$ denote the channel from the BS to all LUEs.
The adopted null-space AN matrix is given by
\begin{equation}
\vect{V}=\vect{I}_\mathrm{N_\mathrm{b}}-\vect{H}_{\mathrm{b}}\left(\vect{H}_{\mathrm{b}}^\mathrm{H}\vect{H}_{\mathrm{b}}\right)^{-1}\vect{H}_{\mathrm{b}}^\mathrm{H}.
\end{equation}
Let $\sigma_\epsilon^2=\epsilon_s^{-1}\big(\mathds{1}_{\mathcal{E}}(u)\epsilon_\mathrm{a}\|\vect{h}_{u\mathrm{b}}^\mathrm{H}\vect{V}\|^2+\sigma^2\big)$,
and let $\vect{W}_{k}=[\vect{w}_{1},\cdots,\vect{w}_{k-1},\vect{w}_{k+1},\cdots,\vect{w}_{K}]$.
The SE for UE-$u$ when decoding the signal of LUE-$k$ is given by \eqref{Rate_origin}, shown at the top of the next page.
\begin{figure*}[!t]
\begin{equation}\label{Rate_origin}
{R}_{u,k}(\vect{W},\epsilon)=\ln\left(1+\frac{\epsilon_\mathrm{s}|\vect{h}_{u\mathrm{b}}^\mathrm{H}\vect{w}_k|^2}
{\epsilon_\mathrm{s}\sum_{j\in\mathcal{K}\setminus{k}}|\vect{h}_{u\mathrm{b}}^\mathrm{H}\vect{w}_j|^2
+\mathds{1}_{\mathcal{E}}(u)\epsilon_\mathrm{a}\sum_{n=1}^{N_\mathrm{b}}|\vect{h}_{u\mathrm{b}}^\mathrm{H}\vect{v}_n|^2+\sigma^2}\right)
=\ln\left(1+\frac{|\vect{h}_{u\mathrm{b}}^\mathrm{H}\vect{w}_k|^2}
{\|\vect{h}_{u\mathrm{b}}^\mathrm{H}\vect{W}_k\|^2
+\sigma_\epsilon^2}\right)
\end{equation}
\hrulefill
\end{figure*}

\subsection{Problem Formulation Assuming Perfect CSI of EUEs}
According to \eqref{Rate_origin}, we define $S_{e,k}(\vect{W},\epsilon)=[R_{k,k}(\vect{W},\epsilon)-R_{e,k}(\vect{W},\epsilon)]^+$ as the SR regarding LUE-$k$ and EUE-$e$.
For the considered system, we aim to maximize the minimum SR, and the optimization problem can be formulated as

\begin{align}\label{Problem_P0}
&\mathcal{P}_0:\max_{\vect{W},\epsilon}\min_{k,e}
\;S_{e,k}(\vect{W},\epsilon)
\\\notag&\text{s.t.}\quad
\mathcal{C}_1:\|\vect{W}\|_\text{F}^2\leq{K},
\\\notag&\qquad\,\mathcal{C}_2:0<\epsilon\leq1,
\end{align}
where $\mathcal{C}_1$ is the transmit power budget, and $\mathcal{C}_2$ defines the allowable range for the power allocation factor.
We first solve the problem $\mathcal{P}_0$ assuming perfect CSI of EUEs. 
Afterwards, we will adapt the proposed solution to formulate and solve a practical problem under no CSI of EUEs.

\section{Proposed Optimization Scheme}

To tackle the non-convex problem in \eqref{Problem_P0}, we alternatively
optimize the beamfocusing matrix and power allocation factor,
which consists of two subproblems.

\subsection{Beamforcusing Matrix Optimization}

For the given power allocation, the original problem $\mathcal{P}_0$ in \eqref{Problem_P0} becomes

\begin{align}\label{Problem_P1}
&\mathcal{P}_1:\max_{\vect{W}}\min_{k,e}
\;S_{e,k}(\vect{W},\epsilon)
\\\notag&\text{s.t.}\quad
\mathcal{C}_1:\|\vect{W}\|_\text{F}^2\leq{K}.
\end{align}
To solve problem $\mathcal{P}_1$, we first introduce the following Lemma.

\begin{lemma}
For $\vect{g}_1\in\mathbb{C}^{1\times{N}_1}$ and $\vect{g}_2\in\mathbb{C}^{1\times{N}_2}$, we have $f_1^{(t)}(\vect{g}_1,\vect{g}_2)\leq\ln\left(1+\frac{\|\vect{g}_1\|^2}{\|\vect{g}_2\|^2+\sigma^2}\right)\leq{f}_2^{(t)}(\vect{g}_1,\vect{g}_2)$,
where
\begin{align}
\notag{f}_1^{(t)}(\vect{g}_1,\vect{g}_2)&=\ln\left(1+\frac{\|\vect{g}_1^{(t)}\|^2}{\|\vect{g}_2^{(t)}\|^2+\sigma^2}\right)
-\frac{\|\vect{g}_1^{(t)}\|^2+\sigma^2}{\|\vect{g}_2^{(t)}\|^2+\sigma^2}
\\\notag&+\frac{\sigma^2}{\|\vect{g}_1^{(t)}\|^2+\|\vect{g}_2^{(t)}\|^2+\sigma^2}+\frac{2\Re[\vect{g}_1(\vect{g}_1^{(t)})^\mathrm{H}]}{\|\vect{g}_2^{(t)}\|^2+\sigma^2}
\\&-\frac{\|\vect{g}_1^{(t)}\|^2(\|\vect{g}_1\|^2+\|\vect{g}_2\|^2)}{(\|\vect{g}_1^{(t)}\|^2+\|\vect{g}_2^{(t)}\|^2+\sigma^2)(\|\vect{g}_2^{(t)}\|^2+\sigma^2)},
\\\notag{f}_2^{(t)}(\vect{g}_1,\vect{g}_2)&=\ln\left(1+\frac{\|\vect{g}_1^{(t)}\|^2}{\|\vect{g}_2^{(t)}\|^2+\sigma^2}\right)
+\frac{\|\vect{g}_2^{(t)}\|^2}{\sigma^2}
\\\notag&-\frac{\|\vect{g}_1^{(t)}\|^2\sigma^2}{(\|\vect{g}_1^{(t)}\|^2+\|\vect{g}_2^{(t)}\|^2+\sigma^2)(\|\vect{g}_2^{(t)}\|^2+\sigma^2)}
\\\notag&+\frac{\|\vect{g}_1\|^2+\|\vect{g}_2\|^2}{\|\vect{g}_1^{(t)}\|^2+\|\vect{g}_2^{(t)}\|^2+\sigma^2}-\frac{2\Re[\vect{g}_2(\vect{g}_2^{(t)})^\mathrm{H}]}{\sigma^2}
\\&+\frac{\|\vect{g}_2^{(t)}\|^2\|\vect{g}_2\|^2}{(\|\vect{g}_2^{(t)}\|^2+\sigma^2)\sigma^2}.
\end{align}
\end{lemma}

\begin{IEEEproof}
See \cite{7447784}.
\end{IEEEproof}

Since the objective function in \eqref{Problem_P1} is non-convex, we adopt the SCA method.
Let $\vect{W}^{(t)}=\big[\vect{w}_1^{(t)},\cdots,\vect{w}_K^{(t)}\big]\in\mathbb{C}^{{N}_\mathrm{i}\times{K}}$ be the feasible solution of $\vect{W}$ at the $t$-th iteration.
The original problem $\mathcal{P}_0$ in \eqref{Problem_P0} can be converted into
\begin{align}\label{Problem_P2}
&\mathcal{P}_2:\max_{\vect{W}}\min_{k,e}
\;{f}_1^{(t)}(\vect{h}_{k\mathrm{b}}^\mathrm{H}\vect{w}_k,\vect{h}_{k\mathrm{b}}^\mathrm{H}\vect{W}_k)
-{f}_2^{(t)}(\vect{h}_{e\mathrm{b}}^\mathrm{H}\vect{w}_k,\vect{h}_{e\mathrm{b}}^\mathrm{H}\vect{W}_k)
\\\notag&\text{s.t.}\quad
\mathcal{C}_1:\|\vect{W}\|_\text{F}^2\leq{K},
\end{align}
where
\begin{align}\label{f1}
\notag&{f}_1^{(t)}(\vect{h}_{k\mathrm{b}}^\mathrm{H}\vect{w}_k,\vect{h}_{k\mathrm{b}}^\mathrm{H}\vect{W}_k)
=R_{k,k}(\vect{W}^{(t)})-\frac{|\vect{h}_{k\mathrm{b}}^\mathrm{H}\vect{w}_k^{(t)}|^2+\sigma_\epsilon^2}{\|\vect{h}_{k\mathrm{b}}^\mathrm{H}\vect{W}_k^{(t)}\|^2+\sigma_\epsilon^2}
\\\notag&+\frac{\sigma_\epsilon^2}{\|\vect{h}_{k\mathrm{b}}^\mathrm{H}\vect{W}^{(t)}\|^2+\sigma_\epsilon^2}+\frac{2\Re[(\vect{h}_{k\mathrm{b}}^\mathrm{H}\vect{w}_k^{(t)})^\mathrm{H}\vect{h}_{k\mathrm{b}}^\mathrm{H}\vect{w}_k]}{\|\vect{h}_{k\mathrm{b}}^\mathrm{H}\vect{W}_k^{(t)}\|^2+\sigma_\epsilon^2}
\\&-\frac{|\vect{h}_{k\mathrm{b}}^\mathrm{H}\vect{w}_k^{(t)}|^2\|\vect{h}_{k\mathrm{b}}^\mathrm{H}\vect{W}\|^2}
{(\|\vect{h}_{k\mathrm{b}}^\mathrm{H}\vect{W}^{(t)}\|^2+\sigma_\epsilon^2)(\|\vect{h}_{k\mathrm{b}}^\mathrm{H}\vect{W}_k^{(t)}\|^2+\sigma_\epsilon^2)},
\end{align}
\begin{align}\label{f2}
\notag&{f}_2^{(t)}(\vect{h}_{e\mathrm{b}}^\mathrm{H}\vect{w}_k,\vect{h}_{e\mathrm{b}}^\mathrm{H}\vect{W}_k)=R_{e,k}(\vect{W}^{(t)})+\frac{\|\vect{h}_{e\mathrm{b}}^\mathrm{H}\vect{W}_k^{(t)}\|^2}{\sigma_\epsilon^2}
\\\notag&-\frac{|\vect{h}_{e\mathrm{b}}^\mathrm{H}\vect{w}_k^{(t)}|^2\sigma_\epsilon^2}
{(\|\vect{h}_{e\mathrm{b}}^\mathrm{H}\vect{W}^{(t)}\|^2+\sigma_\epsilon^2)(\|\vect{h}_{e\mathrm{b}}^\mathrm{H}\vect{W}_k^{(t)}\|^2+\sigma_\epsilon^2)}
\\\notag&+\frac{\|\vect{h}_{e\mathrm{b}}^\mathrm{H}\vect{W}\|^2}{\|\vect{h}_{e\mathrm{b}}^\mathrm{H}\vect{W}^{(t)}\|^2+\sigma_\epsilon^2}
-\frac{2\Re[\tr((\vect{h}_{e\mathrm{b}}^\mathrm{H}\vect{W}_k^{(t)})^\mathrm{H}\vect{h}_{e\mathrm{b}}^\mathrm{H}\vect{W}_k)]}{\sigma_\epsilon^2}
\\&+\frac{\|\vect{h}_{e\mathrm{b}}^\mathrm{H}\vect{W}_k^{(t)}\|^2\|\vect{h}_{e\mathrm{b}}^\mathrm{H}\vect{W}_k\|^2}{(\|\vect{h}_{e\mathrm{b}}^\mathrm{H}\vect{W}_k^{(t)}\|^2+\sigma_\epsilon^2)\sigma_\epsilon^2}.
\end{align}

After introducing the auxiliary variable $\xi$, problem $\mathcal{P}_1$ can be converted into the following problem \cite{10552246}
\begin{align}\label{Problem_P3}
&\mathcal{P}_3:\max_{\vect{W},\xi}
\;\xi
\\\notag&\text{s.t.}\quad
\mathcal{C}_1:\|\vect{W}\|_\text{F}^2\leq{K},
\\\notag&\qquad\,\mathcal{C}_2:{f}_1^{(t)}(\vect{h}_{k\mathrm{b}}^\mathrm{H}\vect{w}_k,\vect{h}_{k\mathrm{b}}^\mathrm{H}\vect{W}_k)
-{f}_2^{(t)}(\vect{h}_{e\mathrm{b}}^\mathrm{H}\vect{w}_k,\vect{h}_{e\mathrm{b}}^\mathrm{H}\vect{W}_k)\geq\xi,
\\\notag&\qquad\,\forall{k}\in\mathcal{K},e\in\mathcal{E},
\end{align}
which can be solved by using CVX.

\subsection{Power Allocation Optimization}

For the given $\vect{W}$, the original problem $\mathcal{P}_0$ in \eqref{Problem_P0} becomes
\begin{align}\label{Problem_P4}
&\mathcal{P}_4:\max_{\epsilon}
\;S_{e,k}(\vect{W},\epsilon)
\\\notag&\text{s.t.}\quad
\mathcal{C}_1:0<\epsilon\leq1.
\end{align}
From \eqref{Rate_origin}, the first-order derivative of $S_{e,k}(\vect{W},\epsilon)$ is given in \eqref{Sek_1st_derivative}, where $A_1=|\vect{h}_{k\mathrm{b}}^\mathrm{H}\vect{w}_k|^2$, $B_1=\|\vect{h}_{k\mathrm{b}}^\mathrm{H}\vect{W}_k\|^2$, $A_2=|\vect{h}_{e\mathrm{b}}^\mathrm{H}\vect{w}_k|^2$, 
$B_2=\|\vect{h}_{e\mathrm{b}}^\mathrm{H}\vect{W}_k\|^2$,
and $V=\|\vect{h}_{u\mathrm{b}}^\mathrm{H}\vect{V}\|^2$.
Since the beamfocusing vector is highly concentrated around the LUE locations, we have $\frac{B_1}{A_1}\rightarrow0$ for $N_\mathrm{b}\gg1$. 
Thus, the optimal $\epsilon$ satisfying $\frac{\partial{S}_{e,k}(\vect{W},\epsilon)}{\partial\epsilon}=0$ is approximated as \eqref{root_epsilon}, where $(a)$ follows from $B_2\ll{A}_2$.
Given $A_2N_\mathrm{b}\gg{K}V$, \eqref{root_epsilon} implies one negative root $\epsilon_{-}^\star$ and one positive root $\epsilon_{+}^\star\in(0,1)$.
Hence, $S_{e,k}(\vect{W},\epsilon)$ is unimodal in $\epsilon$, which means it can be accurately optimizated via the GSS method \cite{9464264}.


\begin{figure*}[!t]
\begin{align}
\notag&\frac{\partial{S}_{e,k}(\vect{W},\epsilon)}{\partial\epsilon}=
KP_\mathrm{b}\bigg[\frac{A_1\sigma^2}{(\sigma^2K+B_1P_\mathrm{b}\epsilon)(\sigma^2K+(A_1+B_1)P_\mathrm{b}\epsilon)}
\\&-\frac{A_2N_\mathrm{b}(\sigma^2N_\mathrm{b}+P_\mathrm{b}V)}{(\sigma^2KN_\mathrm{b}+B_2N_\mathrm{b}P_\mathrm{b}\epsilon
+KP_\mathrm{b}V(1-\epsilon))(\sigma^2KN_\mathrm{b}+(A_2+B_2)N_\mathrm{b}P_\mathrm{b}\epsilon
+KP_\mathrm{b}V(1-\epsilon))}\bigg]\label{Sek_1st_derivative}
\\&\epsilon^\star\approx\frac{-B_2KN_\mathrm{b}V+2K^2V^2\pm\sqrt{-4A_2B_2K^2N_\mathrm{b}^2V^2+B_2^2K^2N_\mathrm{b}^2V^2+4A_2K^3N_\mathrm{b}V^3}}
{2(A_2B_2N_\mathrm{b}^2-(A_2+B_2)KN_\mathrm{b}V+K^2V^2)}
\overset{(a)}\approx\frac{-KV\pm\sqrt{A_2N_\mathrm{b}KV}}{A_2N_\mathrm{b}-KV}\label{root_epsilon}
\end{align}
\hrulefill
\end{figure*}

The full algorithm is presented in Algorithm \ref{Algorithm1}, with $\eta_\text{out}$ as the outer loop precision. 
Since the computational complexity (CC) of GSS is $\mathcal{O}(\ln(1/\eta_\text{gss}))$, where $\eta_\text{gss}$ is the precision, the CC of Algorithm \ref{Algorithm1} is $\mathcal{O}(N_\mathrm{b}^3K^3)$ \cite{10540207}. 
It is noted that the CC of \cite{Zhifeng2025Low} is at the same level as $\mathcal{O}(\ln(1/\eta_\text{gss}))$. However, it considers only the single LUE-EUE case and assumes perfect CSI of EUEs, which imposes significant limitations.

\begin{algorithm}[tp]
\caption{Two-Stage Optimization Scheme}
\label{Algorithm1}
\KwIn{$\{\vect{h}_{u\mathrm{b}}^\mathrm{H}|u\in\mathcal{U}\}$ for $\mathcal{P}_0$ or $\{\vect{h}_{k\mathrm{b}}^\mathrm{H}|k\in\mathcal{K}\}$ for $\mathcal{P}_5$} 
\KwOut{$W^\star$, $\epsilon^\star$}
{\bf{Initialization}}: $\vect{W}^{(0)}=(\vect{H}_{\mathrm{b}}\vect{H}_{\mathrm{b}}^\mathrm{H}+\sigma^2\vect{I}_{N_\mathrm{b}})^{-1}\vect{H}_{\mathrm{b}}$, $\epsilon^{(0)}=1$, $t=1$, $\xi^{(0)}$;
\\\While{$|\xi^{(t)}-\xi^{(t-1)}|/\xi^{(t-1)}<\eta_\emph{\text{out}}$}
{\textbf{\underline{Beamfocusing Matrix Optimization}}
\\For $\forall{k}\in\mathcal{K}$ and $\forall{e}\in\mathcal{E}$, Update $\vect{W}^{(t)}$ by solving problem $\mathcal{P}_3$;

\textbf{\underline{Power Allocation Factor Optimization}}
\\Calculate $\left\{S_{e,k}(\vect{W}^{(t)},\epsilon^{(t-1)})|{k}\in\mathcal{K},{e}\in\mathcal{E}\right\}$ according to \eqref{Rate_origin} and obtain $(e^\star,k^\star)$ which yields the minimum SR;
\\Update $\epsilon^{(t)}$ according to $S_{e^\star,k^\star}(\vect{W}^{(t)},\epsilon^{(t-1)})$;
\\$t=t+1$;}
\bf{Return}: $\vect{W}^\star=\vect{W}^{(t)}$, $\epsilon^\star=\epsilon^{(t)}$.
\\\bf{final};
\end{algorithm}

\begin{remark}
Under the condition of unknown CSI of EUEs, the original optimization problem is written as
\begin{align}\label{Problem_P5}
&\mathcal{P}_5:\max_{\vect{W},\epsilon}\min_{k,e}
\;R_{k,k}(\vect{W},\epsilon)
\\\notag&\text{s.t.}\quad
\mathcal{C}_1:\|\vect{W}\|_\text{F}^2\leq{K},
\\\notag&\qquad\,\mathcal{C}_2:0<\epsilon\leq1,
\end{align}
which can be solved by using the same method as Algorithm \ref{Algorithm1} by setting ${R}_{e,k}(\vect{W})$ to zero.
\end{remark}

\section{Simulation Results and Discussion}

This section presents the simulation results for the considered NF communication system. The cell size is $100\times100\:\text{m}^2$.
The BS is located at the coordinates $(50,100)$ m, and the EUEs are positioned in the same direction as the LUEs relative to the BS.
Specifically, we consider a urban microcell model with a $2$ GHz carrier frequency, and the LSFC given as
\begin{equation}
\beta_{u\mathrm{i}}[dB]=-33.05-36.7\log_{10}(d_{u\mathrm{i}}).
\end{equation}
In addition, assuming a $20$ MHz transmission bandwidth and a $5$ dB noise figure, the noise power $\sigma^2$ is set to $-96$ dBm \cite{8845768}.
Typically, we consider two different CSI conditions:
\begin{itemize}
  \item $\mathrm{S}_1$: Unknown CSI of EUEs, where $\mathcal{P}_5$ is solved.
  \item $\mathrm{S}_2$: Perfect CSI of EUEs, where $\mathcal{P}_0$ is solved. The SR obtained based on $\mathrm{S}_2$ acts as the upper bound of that obtained based on $\mathrm{S}_1$.
\end{itemize}
Additionally, four beamforming/beamfocusing schemes are considered:
\begin{itemize}
  \item Proposed scheme: Signal beamfocusing based on Algorithm \ref{Algorithm1} with AN.
  \item Scheme in \cite{10540207}: Signal beamfocusing without AN.
  \item Scheme in \cite{Zhifeng2025Low}: MRT beamfocusing with AN.
  \item FF beamforming (FFB) based on the FF channel model.
\end{itemize}

\subsection{SR with Power Allocation}

\begin{figure}
\centering
\subfigure[]{
\begin{minipage}[t]{0.91\linewidth}
\includegraphics[width=1\linewidth]{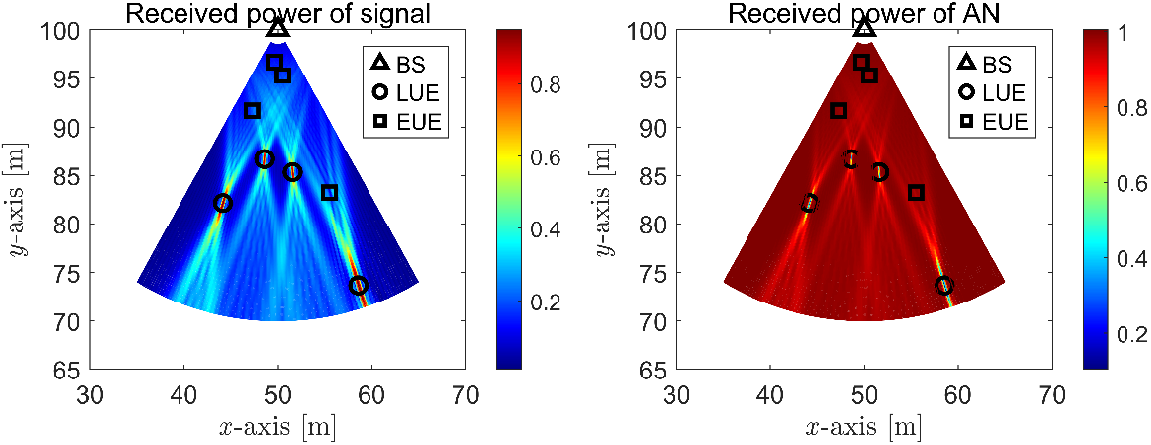}
\end{minipage}
}

\subfigure[]{
\begin{minipage}[t]{0.9\linewidth}
\includegraphics[width=1\linewidth]{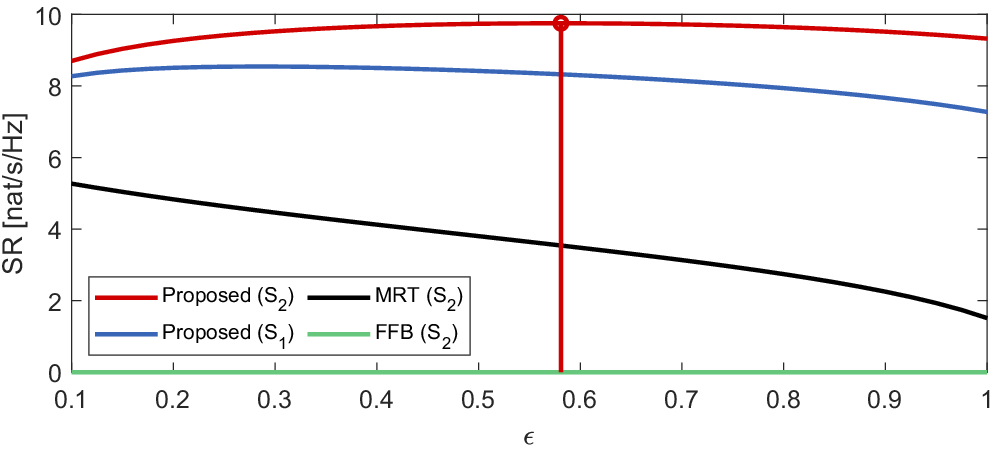}
\end{minipage}
}
\vspace{-0.1 in}
\caption{Secrecy performance of the considered XL-MIMO. (a) Beam patterns of the received power of signal and AN. (b) Minimum SR versus power allocation $\epsilon$.}
\label{SR_power_allocation}
\end{figure}

Fig. \ref{SR_power_allocation} compares the minimum SR of different  beamforming/beamfocusing schemes in the XL-MIMO system, where the power allocation factor $\epsilon$ ranges from $0.1$ to $1$.
For the multiple LUE-EUE case, we set $K=E=4$ and $P=0$ dBm.
Fig. \ref{SR_power_allocation}(b) shows that FFB achieves negligible SR compared to NF beamfocusing (NFB), indicating that angular domain-based beamfocusing design is ineffective in preventing wiretapping in the NF scenario.
The reason for this is explained in Fig. \ref{SR_power_allocation}(a). Unlike the FF signal beam pattern, where LUEs and EUEs in the same direction experience nearly equal channel gain \cite{9903389}, the NF beam pattern provides high gain only to LUEs within the focused area.
Additionally, Fig. \ref{SR_power_allocation}(a) shows that the null-space AN and signal beam patterns are complementary, with AN covering the entire NF region except for the focused area of LUEs.
This suggests that generating null-space AN can further reduce the wiretapping capacity in the NF scenario.

Moreover, Fig. \ref{SR_power_allocation}(b) demonstrates that the proposed two-stage beamfocusing significantly enhances SR compared to MRT beamfocusing, which performs best at $\epsilon\approx0.58$.
Finally, by comparing the unknown and perfect CSI cases of EUEs ($\mathrm{S}_1$ and $\mathrm{S}_2$), it can be observed that the SR obtained by solving $\mathcal{P}_0$ serves as an upper bound for that obtained by solving $\mathcal{P}_5$.

\subsection{SR with Transmit Power Budget}

\begin{figure}
\begin{center}
\includegraphics[width=0.9\columnwidth]{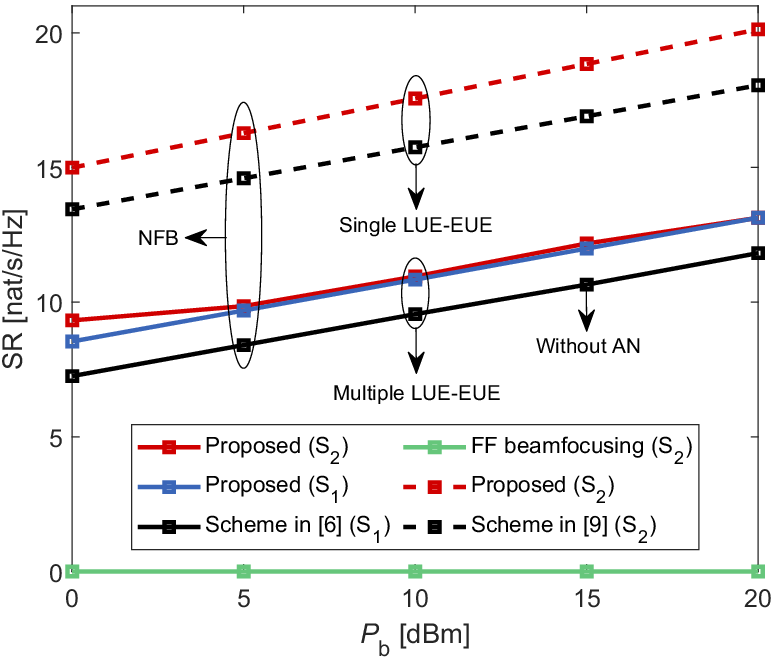} 
\captionsetup{font={normal}}
\caption{Minimum SR versus transmit power budget $P_\mathrm{b}$ for the proposed beamfocusing matrix.}
\label{SR_power_budget}
\end{center} \vskip-3mm
\end{figure}

Fig. \ref{SR_power_budget} compares the minimum SR of different  beamforming/beamfocusing schemes in the XL-MIMO system, where the transmit power budget $P_\mathrm{b}$ ranges from $0$ dBm to $20$ dBm.
For the multiple LUE-EUE case, we set $K=E=4$.
Consistent with Fig. \ref{SR_power_allocation}(b), SR with FFB remains negligible, unaffected by $P_\mathrm{b}$ variations.
Moreover, the SR of the proposed NFB scheme under perfect CSI ($\mathrm{S}_2$) forms a tight upper bound for that under unknown CSI ($\mathrm{S}_1$), with the gap narrowing as $P_\mathrm{b}$ increases.
Furthermore, in the unknown CSI ($\mathrm{S}_1$) case, the proposed NFB scheme outperforms the no-AN transmission in \cite{10540207}, underscoring the role of AN injection in enhancing SR in NF communications.

For the single LUE-EUE case, we follow the setup in \cite{Zhifeng2025Low} with $K=E=1$. 
In contrast to the multiple LUE-EUE case, the SR of single LUE-EUE case achieves higher value due to the absence of multi-UE interference.
Moreover, the proposed NFB scheme outperforms the adopted scheme in \cite{Zhifeng2025Low}, demonstrating its superiority in both scenarios.

\section{Conclusion}

This paper investigated secure DL transmission in an NF XL-MIMO system, focusing on enhancing the SR through the transmission of null-space AN alongside the confidential message. By optimizing the NF beamfocusing matrix and power allocation under various CSI conditions, the proposed two-stage algorithm significantly improved SR performance.
Key findings include the effective generation of highly focused and complementary beam patterns in NF, which substantially outperformed conventional FFB. Additionally, the proposed beamfocusing scheme demonstrated superior SR compared to conventional MRT beamfocusing, achieving the highest SR with optimal power allocation. Furthermore, as transmit power increased, the performance gap between perfect and unknown CSI conditions for EUEs narrowed, suggesting that perfect CSI is a tight upper bound in high-SNR scenarios.
These results highlight the potential of NF XL-MIMO systems for achieving enhanced secure communication in practical wireless networks..

\bibliography{IEEEabrv,journal}
\bibliographystyle{IEEEtran}

\end{document}